\documentstyle[iopconf1,psfig]{article}
\begin{document}

\newcommand{\beq}{\begin{eqnarray}}
\newcommand{\eeq}{\end{eqnarray}}
\newcommand{\eqna}{\begin{eqnarray}}
\newcommand{\eqne}{\end{eqnarray}}
\newcommand{\dia}{\begin{displaymath}}
\newcommand{\die}{\end{displaymath}}
\newcommand{\eqnaa}{\begin{eqnarray*}}
\newcommand{\eqnae}{\end{eqnarray*}}
\def\dleft{\rlap{{\it D}}\raise 8pt\hbox{$\scriptscriptstyle\Leftarrow$}}
\def\prop{\propto}
\def\dright{\rlap{{\it D}}
\raise 8pt\hbox{$\scriptscriptstyle\Rightarrow$}}
\def\lrartop#1{#1\llap{
\raise 8pt\hbox{$\scriptscriptstyle\leftrightarrow$}}}
\def\_#1{_{\scriptscriptstyle #1}}
\def\&#1{^{\scriptscriptstyle #1}}
\def\sss{\scriptscriptstyle}
\def\dij{\delta_{\sss ij}}
\def\abs#1{\vert #1\vert}
\def\rearth{R_{\oplus}}
\def\mearth{M_{\oplus}}
\def\tearth{t_{\oplus}}
\def\hearth{H_{\oplus}}
\def\cearth{c_{\oplus}}
\def\d{\delta}
\def\gij{g_{\sss ij}}
\def\Gij{g^{\sss ij}}
\def\gkm{g_{\sss km}}
\def\Gkm{g^{\sss km}}
\def\cd#1{{}_{\sss;#1}}
\def\ud#1{{}_{\sss,#1}}
\def\udu#1{{}^{\sss,#1}}
\def\upcd#1{{}_{\sss;}{}^{\sss #1}}
\def\upud#1{{}_{\sss,}{}^{\sss #1}}
\def\ro{r\_{0}}
\def\vro{{\bf r}\_{0}}
\def\vr{{\bf r}}
\def\vv{{\bf v}}
\def\vg{{\bf g}}
\def\vgn{{\bf g}_{\rm N}}
\def\gn{{\rm g_N}}
\def\ve{{\bf e}}
\def\vx{{\bf x}}
\def\rar{\rightarrow}
\def\deriv#1#2{{d#1\over d#2}}
\def\oot{{1\over 2}}
\def\pdline#1#2{\partial#1/\partial#2}
\def\pd#1#2{{\partial#1\over \partial#2}}
\def\pdd#1#2#3{{\partial^2#1\over\partial#2\partial#3}}
\def\av#1{\langle#1\rangle}
\def\avlar#1{\big\langle#1\big\rangle}
\def\div{{\vec\nabla}\cdot}
\def\grad{{\vec\nabla}}
\def\curl{{\vec\nabla}\times}
\def\DD{{\cal D}}
\def\m{\mu}
\def\n{\nu}
\def\eps{\epsilon}
\def\vq{{\bf q}}
\def\dtv{d\&3v}
\def\f{\varphi}
\def\fb{\f\&{B}}
\def\eq{E_q}
\def\fl{\varphi}
\def\gf{\grad\f}
\def\b{\beta}
\def\c{\gamma}
\def\l{\lambda}
\def\L{\Lambda}
\def\om{\omega}
\def\bk{\par\noindent}
\def\a0{a_0}
\def\ao{a_0}
\def\h0{H_0}
\def\tl{T\_{\L}}
\def\va{{\bf a}}
\def\vA{{\bf A}}
\def\vD{{\bf D}}
\def\vR{{\bf R}}
\def\vF{{\bf F}}
\def\cmss{cm~s^{-2}}
\title{The Modified Dynamics--A Status Review }

\author{Mordehai Milgrom
\footnote{E-mail:fnmilgrm@wicc.weizmann.ac.il}}

\affil{Department of Condensed-Matter Physics, Weizmann Institute, Rehovot 76100, Israel}

\beginabstract
The Modified dynamics (MOND) has been propounded as an alternative to
Dark matter. It imputes the mass discrepancy in galaxy systems 
to failure of standard dynamics in the limit of small accelerations.
After a brief description of the MOND tenets, I discuss how its predictions
now compare with the data. I put special emphasis
on rotation-curve analysis--whence  comes the most clear-cut support
for MOND, and on the cores of rich x-ray clusters, where MOND
does not yet explain away the mass discrepancy. I then outline the 
MOND program, especially work still left to do.
This is followed by general comments on cosmology and structure formation
in MOND. I conclude with some incomplete thoughts on the possible origin
of MOND (as an effective theory); in particular on the possibility
that it comes about as a vacuum effect.
\endabstract

\section{Introduction--the basic tenets of MOND\label{introduction}}
To speak of the ``dark matter'' problem is to beg one of the most important
 conundrums in present-day science; after all  we have no direct 
evidence that dark matter actually exists in appreciable quantities.
 All we know is that the masses directly observed in galactic systems fall below
 what is calculated using standard dynamics. Stuffing galactic systems
and the universe with putative dark matter is perhaps the least
painful remedy for most people,
 but it is not the only one possible. Another avenue
 worthy of consideration builds on a possible failure of standard dynamics
under the conditions that prevail in galactic systems. 
As you may know, the modified dynamics (MOND) has been put forth in just this
 vein\cite{mond}.
 It hinges on the accelerations in galactic
 systems being very small compared with what is encountered in the solar system,
say.
 MOND asserts that 
non-relativistic dynamics involves the constant
$\a0$, with the dimensions of acceleration, so that in the formal limit
$\a0\rar 0$--i.e., when all quantities with the dimensions of acceleration are
 much larger
than $\a0$--standard dynamics obtains (in analogy with the appearance
 of $\hbar$ in 
quantum mechanics, and the classical limit for $\hbar\rar 0$).
 In the opposite (MOND) limit of large
$\a0$ dynamics is marked by reduced inertia;
 one may roughly say that in this limit inertia at acceleration $a$ is
 $ma^2/\a0$, instead of the standard $ma$.
 This still allows for different
specific formulations. Indeed
 we have nonrelativistic
formulations of MOND, derivable from actions, based on either modified 
gravity\cite{bm}, or on modified inertia\cite{ann}; these will be described
 below. 
 A simple, if primitive, formulation that captures 
much of the content of MOND, and which gives the basic idea, is this:
 Imagine a test particle in the gravitational field 
of some mass distribution whose standard (Newtonian) gravitational acceleration
 field is $\vgn$. In standard dynamics the acceleration, $\vg$,
 of the particle is
 $\vgn$ itself. MOND posits that this is so only in the limit 
 $\gn\gg\a0$. In the
 opposite limit $\gn\ll\a0$ we have roughly ${\rm g}\sim (\gn\a0)^{1/2}$.
 To interpolate between the 
limits we use a relation of the form $\mu({\rm g}/\a0)\vg=\vgn$,
 where $\mu(x)\approx x$ for $x\ll 1$, and
$\mu(x)\approx 1$ when $x\gg 1$.
 This relation gives an approximate relation between the typical
 accelerations
in a system (as embodied, say, in an exact virial relation derived from an
exact theory). It also gives a very good approximation for the acceleration
 in circular motion relevant for rotation curves of disc 
galaxies\cite{ann}\cite{sol} (in modified inertia theories it give the exact rotation curve).
 In the 
 more decent formulations of MOND, the actual acceleration of
 a test particle is
 not directly related to the local Newtonian acceleration as in the above 
relation (in particular, the two are not in the same direction,
 in general). 
\par
Some immediate, and unavoidable, predictions of even the basic tenets are\cite{mond}\cite{galaxies}:

1. The rotation curve for any isolated body becomes flat, asymptotically.

2. The asymptotic rotational velocity, $V_{\infty}$, depends only on the total mass of the body,
 $M$, via $V^4_{\infty}=MG\a0$. This predicts a Tully-Fisher relation between velocity and luminosity
if the $M/L$ values are narrowly distributed.

3. A similar approximate relation exists,
 for a body supported by random motions,
 between the mean velocity dispersion and the total mass.
 This is relevant to mass determinations of systems such as
dwarf-spheroidal, and elliptical, galaxies, and of galaxy
 groups and clusters. It also predicts an
 approximate $L\propto \sigma^4$ relation in such systems
 (with similar $M/L$ values).

4. The smaller the typical acceleration of a gravitationally
 bound system, the larger the 
 mass discrepancy it should evince. It had thus
 been predicted that all 
low-surface-brightness (LSB) systems should evince large mass
 discrepancies since, for a given
$M/L$, surface brightness is proportional to acceleration
 (in the mean). This pertains, e.g., to
dwarf-spheroidal satellites of the Milky Way, and to
 low-surface-brightness disc galaxies.

5. Above all, the full rotation curve of a disc galaxy
 should be obtained, using MOND,
 from the distribution of the observed mass alone. 
\par
Comparison with the data, as discussed later, yields
 a value of $\a0$, determined in several,
independent ways (using the different roles of $\a0$ in the theory).
Very interestingly, the value $\a0$ turns out to be of the
 same order as $c\h0$--an acceleration
parameter of cosmological significance\cite{mond}. Anticipating
 later discussion, I remark here that 
this might be a crucial clue as to the origin of MOND, and its possible
 origin in effects related to cosmology.

\section{The performance of MOND\label{data}}
\begin{figure}
\psfig{figure=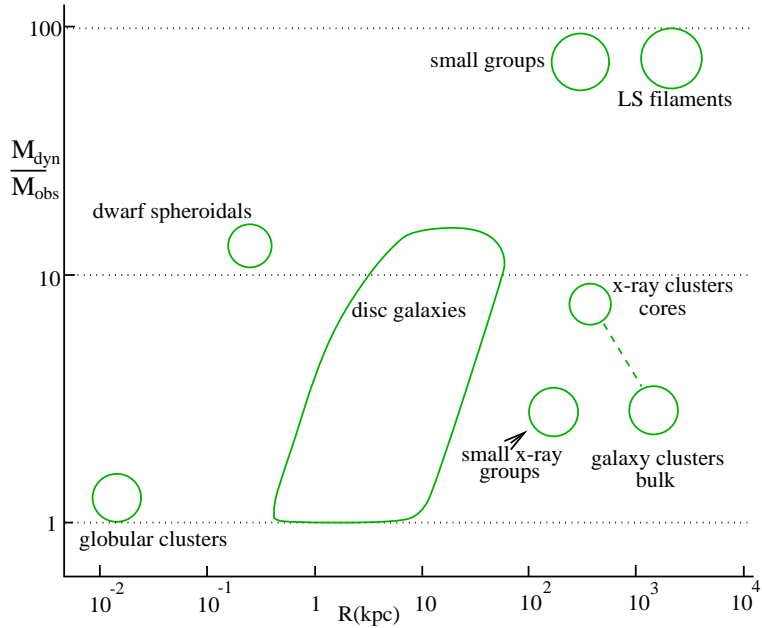,width=10cm}
\caption{The mass discrepancy (dynamical mass over detected mass)
 in various
 galactic systems plotted against the typical system size.}
\end{figure}
Figure 1 summarizes the  mass discrepancy in various galactic
 systems. It shows, approximately, the
ratio of the dynamical mass, as determine with standard
dynamics, to the mass so far accounted for by direct observations.  
The discrepancy is plotted against some ``typical'' system radius.
(Masses in galactic systems show no sign of saturation
with radius, and the value within that ``typical'' radius is used.) 
I note, in passing, that there is no correlation of
 the discrepancy with system size. 
Remark, in particular, that  the small dwarf spheroidals and LSB discs
show large discrepancies, while the large galaxy clusters evince only
 moderate discrepancies. This flies in the face of attempts
 to explain away the mass discrepancy by modifying gravity at
 large distances, predicting 
 increase in the ``discrepancy'' with size. (Contrary to some lingering
misconception, MOND is not a modification at large distances, but at low
accelerations--which for a given mass are attained at large distances.)

The use of MOND dynamics should eliminate the mass discrepancy
 in all systems.  
Put differently, MOND  predicts the mass discrepancy
expected  when using Newtonian dynamics. 
\begin{figure}
\psfig{figure=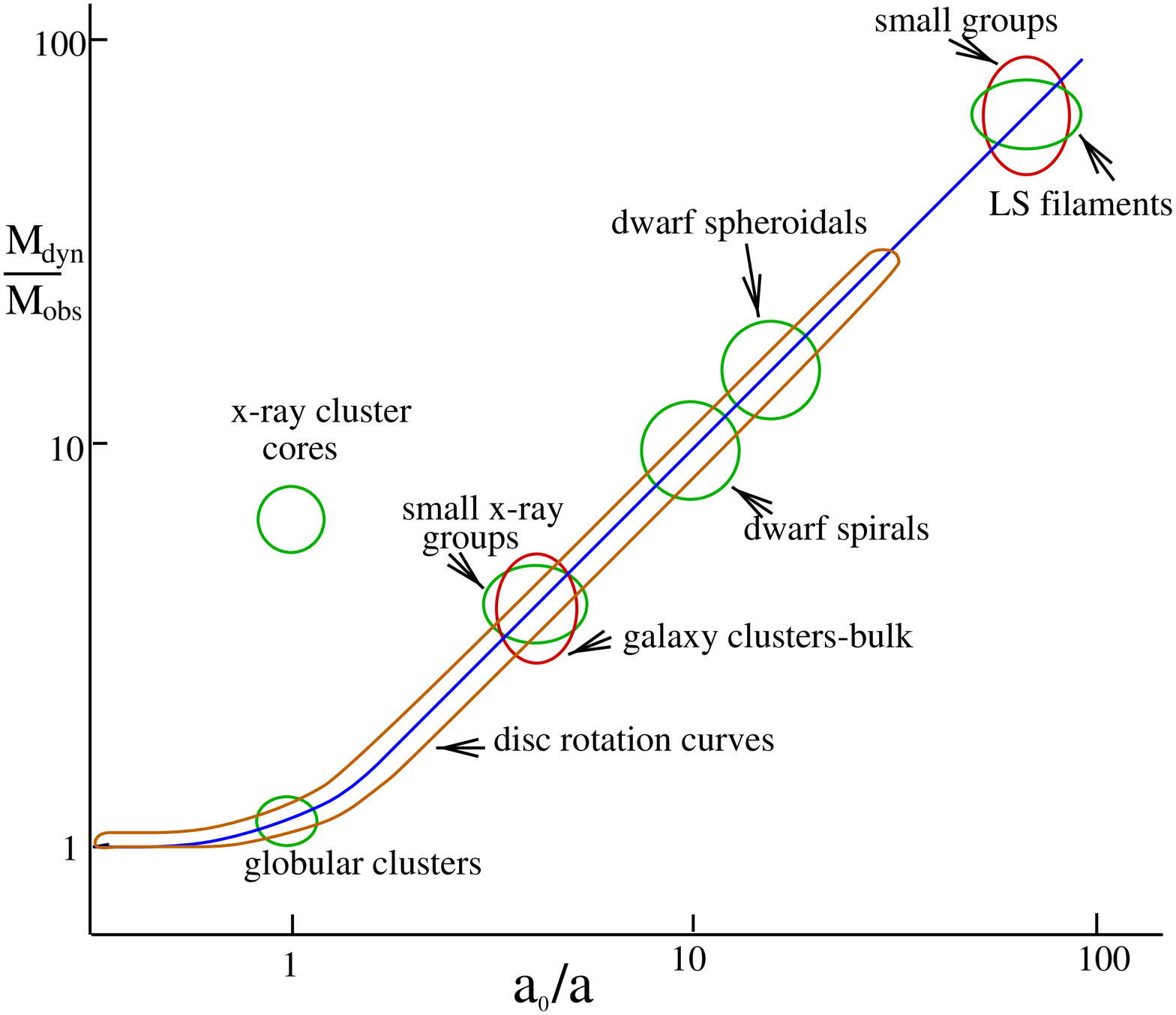,width=10cm}
\caption{The mass discrepancy plotted against the typical
 system acceleration.}
\end{figure}
Figure 2 shows the
discrepancy plotted now against the typical inverse
acceleration--as prescribed by MOND. It also shows the MOND prediction
of the discrepancy as a
 solid line interpolating the value 1 at low $a^{-1}$ and the
 predicted discrepancy, $\a0/a$, at $a\ll \a0$.
The positions of the blobs describing the different galactic systems 
roughly represent detailed work on individual systems:
 dwarf spheroidals\cite{gerhard}-\cite{mateo},
disc-galaxy rotation curves\cite{bbs}-\cite{demc},
 galaxy groups\cite{groups}, 
 x-ray clusters (e.g. \cite{thewhite}-\cite{sanclustb}),
 and  large-scale filaments\cite{filaments}.
I expand here on two types of systems.

\subsection{Cores of rich x-ray clusters}
We see in Fig. 2 that MOND explains away the mass discrepancy in all the systems studied except
 one--the cores of  rich x-ray clusters.
This is discussed in \cite{rome}, but the point had been made before
 in ref.\cite{gerbal} (and  a hint of it is in ref.\cite{thewhite}, see
 also ref.\cite{sanclustb}).
(I separate, somewhat artificially, 
the results for cluster cores, within  a few hundred kiloparsecs, from the
 cluster bulk within a few megaparsecs. There is, of course, continuity:
the discrepancy in the core, which still lingers in MOND, decreases and
 disappears as we go to larger radii.)
 The point is that x-ray-cluster cores have, by and large,  borderline 
 accelerations  (i.e. of order $\a0$ or somewhat larger).
MOND tells us then not to expect much of a  mass discrepancy there,
 when, in fact, the mass so far accounted for (in hot gas, and stars)
falls short of the dynamical mass obtained from gas hydrostatics,
 and from strong lensing.
According to MOND there must then reside in these cores
 normal baryonic matter
 yet undiscovered. It is well known that such clusters
 are characterized by cooling flows
that deposit large quantities of matter in their cores.
 These deposits have not yet been discovered,
and, it is surmised, might be in the form of dim stars or warm gas.
Present-day mass-deposition rates do not suffice to supply
 the required mass within the Hubble time,
but the rates might have been higher in the past. In any
 event it is a strong
 prediction of MOND that the dark matter in cluster cores
 is baryonic and will be detected.
 The recent detection of strong UV emission from the
 cluster Abell 1795 has
 been interpreted as arising from warm gas enough to
 account for the dark matter in the core\cite{mittaz}.
\par
 From an historical perspective, it is interesting to
 remember that at the time of
 the advent of MOND it was not known that clusters harbor
 large quantities of
 hot, x-ray-emitting gas. This, as we now know, constitutes
 the lion's share of the baryonic mass in x-ray clusters.
Similar to the case with the cluster cores now, the MOND analysis of
 the time\cite{systems}
still left a mass discrepancy for some clusters (such as 
Coma, A2029, A2199, A2256).  Seeing that these clusters are x-ray sources,
 it had been surmised\cite{systems}
 that intergalactic gas responsible for the emission might account 
for the lingering discrepancy, as indeed proved to be the case.

\subsection{Rotation-curve analysis}
Rotation-curve analysis is arguably the heart of MOND testing.
 It surpasses all other tests
as regards the quality of the data, the freedom from astrophysical assumptions,
 and the range of acceleration values covered.
About eighty disc galaxies with sufficient data (extended, two-dimensional
 velocity maps, photometry, and HI distribution) have now been successfully
 MOND tested by various
 studies\cite{kent}\cite{bbs}\cite{morishima}\cite{sanrot}\cite{sanver}\cite{demc}.
For each galaxy the analysis involves, in most cases, one adjustable
 parameter--the $M/L$ value of the stellar disc (in standard dark-halo
 fits there are two additional free parameters characterizing the halo).
 A success of MOND for even a single rotation curve is most significant, because  
even full freedom to choose $M/L$ is  anything but sufficient to make MOND
 work in any given case.
This is nicely demonstrated in ref \cite{demc} by analyzing a synthetic galaxy
 model taken to have the HI data (HI distribution, and rotation curve)
from one galaxy, and the stellar light distribution from another.
 An attempt to fit this galaxy with MOND
gives a very bad best fit, and the best-fit $M/L$ value is unreasonably high.
  In contrast, a standard, dark-halo fit for this ``wrong''
galaxy model gives a very good fit (with reasonable $M/L$ value).
Another example serving to demonstrate the limited leverage of the $M/L$ parameter:
Many galaxies have high accelerations ($a\gg\a0$) in their inner parts; 
MOND then predicts no discrepancy there, and the stellar
 $M/L$ value is thus fixed by the inner parts.
The rotation curve in the outer parts (its shape, whether falling or rising,
 and amplitude)
then remains an unadjustable prediction of MOND that could easily fail.
\par
In addition, note that the stellar $M/L$ value is not really a totally
 free parameter.
 In must fall within some acceptable range,
and, by and large,  is constrained by theoretical models. 
The study of ref.\cite{sanver}, which is unique in its use of the infrared $K'$
 photometric band--arguably the best representative  of  stellar mass--shows
 that, indeed, the resulting MOND $M/L$ values, for the sample of
 Ursa Major galaxies studied, are very narrowly distributed near one solar unit.
The study also finds that the B-band, MOND $M/L$ values are 
strongly correlated
 with the observed galaxy color, following the expected theoretical 
relation.  All this shows $M/L$ to be a rather tightly tethered parameter,
 which further heightens the significance of the successful MOND analysis.
\section{The general MOND program\label{program}}
\begin{figure}
\centerline{\psfig{figure=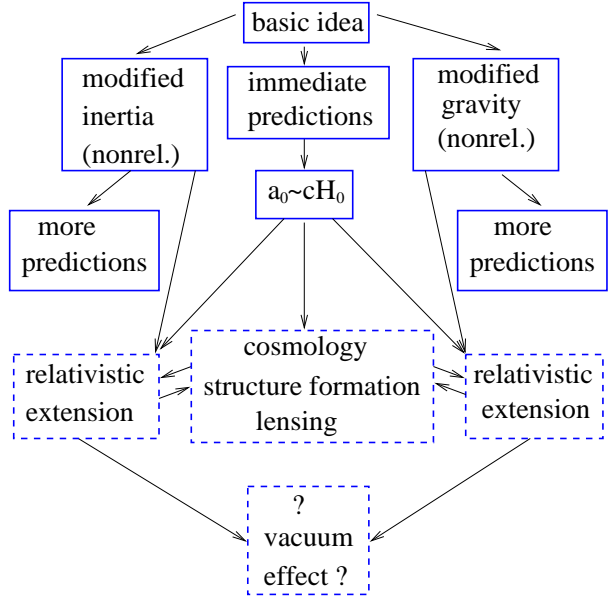,width=8cm}}
\caption{Schematics of the MOND program.}
\end{figure}
Those  cleaving to Newtonian dynamics may take the
 success of MOND to reflect some very strict regularity--encompassing
 the whole gamut of galactic systems--relating
 the distribution of visible matter to that of 
dark matter via a simple formula. The few of us who have contributed
 to MOND theory and testing over the years view this success
as strong indication of departure from standard dynamics
 in the parameter region relevant to galactic systems.
Taking MOND in such a vein, one seeks to
 construct theories, with increasing
 depth and compass, that incorporate the basic tenets of MOND. 
Figure 3 presents the schematics of these efforts, with
 full-line blocks marking areas in advanced stages of development.
\par
 At the nonrelativistic level, at least, MOND may be viewed as either a modification
of gravity, or a modification of inertia\cite{mond}.
In the former, the gravitational field produced by a given mass distribution
is dictated by a new equation; in the latter the equation of motion is 
MONDified, while the force fields remain intact.
An example of the former is the MONDification of the Poisson equation
discussed in ref. \cite{bm} where the gravitational potential, $\f$, is
determined by the mass distribution, $\rho$, via
\beq \div[\mu(\abs{\gf}/\a0)\gf]=4\pi G\rho. \label{ppp}\eeq
\par
Mondified inertia is discussed in refs. \cite{ann}\cite{vacuum}. In such 
 theories, when derived
from an action, one replaces the standard kinetic action for a particle
($\int v^2/2~dt$) by a kinetic action that is a more complicated functional
of the particle trajectory 
 \beq A_mS[\vr(t),\a0], \label{gtarata} \eeq
where $A_m$ depends only on the body, and can be identified with its mass,
and $S$ depends only on the trajectory and on $\a0$ as a parameter.
Weak equivalence is thus insured. In the formal limit $\a0\rar 0$
the action goes to the standard kinetic action.
In the opposite limit, $\a0\rar\infty$, $S\propto \a0^{-1}$,
 and inertia disappears in the very limit.
\par
 With respect to Newtonian dynamics, special relativistic dynamics is an
 example of modified inertia:
The equation of motion of  a relativistic particle
moving in a force field $\vF(\vr)$  is
$md(\c\vv)/dt=m\c[\va+\c^2c^{-2}(\vv\cdot\va)\vv]=\vF(\vr)$,
derived from the kinetic action
$mc^2\int d\tau=mc^2\int\gamma^{-1}~dt$.
 Here too there appears a parameter, $c$, which, like
$\a0$ in MOND (and $\hbar$ in QM) both delimits the standard (classical)
region, and enters the dynamics in the non-classical regime. 
Unlike the special-relativistic action, which is still local, the MOND action
is perforce non-local if it is to be Galilei invariant\cite{ann}.
\par
Mondified gravity and mondified inertia do not differ on what we call
 the basic predictions of MOND:
The asymptotic flatness of rotation curves (and their general shape),
the $M\propto V^4$ relation, the added stability of systems in the deep MOND
 regime\cite{stab}, etc. There are, however, important differences;
 some examples are:
1. In mondified gravity only systems governed by pure gravity (such as 
galactic systems) are affected, while in mondified inertia the
modification applies for whatever combination of forces is at play.
2. in the former, the acceleration of a test particle depends only 
on its position in the field, while in the latter it depends strongly on
 other
details of the trajectory (inertia is identified with acceleration only in
standard Newtonian dynamics).
 As an example, we can see in the 
special-relativity case, mentioned above,
 that the $\vv\cdot\va$ term vanishes for a circular 
orbit, but dominates, at high $\gamma$, for a linear trajectory.  
3. In mondified inertia the expressions for the conserved quantities and  
adiabatic invariants in terms of the motion are modified\cite{ann}, in
 contradistinction to mondified gravity.   
\par
 An acceptable relativistic extension for MOND is not yet at hand.
 Discussions of various candidates can be found in refs.
\cite{bm}\cite{bekrel}\cite{bsbending}\cite{stratified}, but each of these 
has its problems. These problems  seem to be specific to the particular
 models (e.g. that in \cite{bm} has superluminal modes, scalar-tensor
theories as discussed in \cite{bsbending} do not give as large a light
bending as is observed, and that in \cite{stratified} is based on 
a non-dynamical pregeometry).
\par
Reflection over this question has convinced me that a relativistic extension
 will not just be a relativistic theory where $\a0$ appears as a parameter,
with GR restored in the limit $\a0\rar 0$. I have always viewed MOND as an 
effective theory (i.e. an approximate theory that results from a deeper one
in a certain limit, and/or when some of the relevant degrees of freedom
are integrated out). In the present case MOND is perhaps an approximation
in the limit of small sizes and short times (on the cosmological scale), and nonrelativistic motion, due to some yet-undiscovered effect connected with 
cosmology.
An analogy will highlight the point: If we are ignorant of earth gravity
as derived from the pull of the earth--such as when we are immured forever
in a small laboratory near the earth's surface--dynamics is described
approximately by modified inertia of the form
\beq \vF=m(\va-\vg), \label{hutara} \eeq
where $\vF$ is the applied force excluding earth gravity, and $\vg$ is the 
free-fall acceleration on earth.
This can be recast to resemble MOND inertia:
\beq \vF=m\lrartop{\mu}(\va/{\rm g})\cdot\va,  \label{kutatia} \eeq
where
 $\lrartop{\mu}(\vx)\equiv1-{\ve\otimes\vx \over {\rm x}^2}$, and
$\ve\equiv \vg/{\rm g}$ is a down-pointing unit vector.
This is a good approximation inasmuch as this proverbial
 laboratory is our whole universe; i.e.,
 for systems small compared with $\rearth$ (analogous to
the Hubble distance), and times small compared with
 $\hearth^{-1}=\tearth\equiv\rearth/\cearth$,
where $\cearth=(\mearth G/\rearth)^{1/2}$ is the escape speed--analogous
 to the speed of light.
 The effective ``acceleration constant'', g, appearing
 in this modified inertia is 
 related to the ``cosmological'' parameters by
 ${\rm g}=\cearth\hearth$.  

\par
 In a relativistic extension of MOND, or in the cosmological
 context, $\a0$ may lose its role as a
 ``universal constant''\cite{ann}\cite{vacuum}--as g does in the above
 analogy when dealing with, say, satellite motion for which 
$v\sim\cearth$. The peculiar situation is further
highlighted by the fact that--in view of $\a0\sim c\h0$--the
 only system that is both high-field in the GR sense
 and in the deep-MOND regime is the universe at large.
 (In the quantum analogue a system in the high-field, quantum regime is of 
Planck scale or smaller. There, we can, at least look from outside the Planck
scale, which we cannot do in MOND.) Relativistic
MOND must then be understood as part and parcel of cosmology, as I 
elaborate more in the next section.

\section{Cosmology and structure formation\label{cosmology}}
 Cosmology is then not simply an application of a relativistic
 version of MOND but a unit with it.
 The key to finding the underlying theory
 may lie in understanding first how an acceleration
of cosmological significance can, at all, enter local dynamics, which I
 discuss in the last section. 
\par
 If $\a0$ is a fingerprint of
 cosmology on local dynamics, it is not necessarily the identification
$\a0\sim a_{ex}\equiv cH_0$ which is the the right one.
 There are other cosmological acceleration scales\cite{com}\cite{ann}\cite{sivaram}
 such as $a_c\equiv c^2/R_c$, where $R_c$ is the
 curvature radius (spatial
or space-time), or $a\_{\L}\equiv c\L^{1/2}$, where $\L$ is the 
 cosmological constant. Today we have only upper limits on $a_c$, which 
 is of the order of $a_{ex}$. Several pieces of evidence seem now to
imply a non-zero cosmological-constant with 
$\L\sim \h0^2$. If this is true then we also have   $\a0\sim a\_{\L}$.
Thus $\a0$ might be a proxy for any of the cosmological acceleration
parameters.
Since these depend differently on cosmic time, $\a0$ may vary
with cosmic time  in a way that is difficult to know without
the correct identification. Such possible variation of $\a0$ has 
obvious ramifications for the formation and the ensuing evolution
of galactic systems.  
\par
Even without a theory we can make out some semi-quantitative aspects by which
MOND cosmology must differ greatly from standard cosmology:
\par
1. MOND is based on the phenomenology of galactic systems and hence, in 
principle, is not committed on the question of cosmologically
homogeneous component of dark matter. But certainly, it is in the spirit
of MOND that we should not conjecture the existence of any DM component
 without first trying to 
explain it away with new physics. Recent leanings toward a non-zero
cosmological constant(CC) are a step in this direction. And
 perhaps\cite{vacuum} the same mechanism that produces a  
CC-like contribution might also effect MOND (hence the 
coincidence $\a0\sim c\L^{1/2}$). At any event, a MOND-inspired cosmology
would start with no dark matter.
\par
2. The MOND Jeans mass--a basic concept in structure formation, which 
indicates which masses are likely to collapse from an homogeneous
medium--depends differently on the temperature, $T$,
and density, $\rho$, of the medium\cite{com}:
$M_J({\rm MOND})\propto T^2/\a0$, instead of the Newtonian dependence
 $M_J\propto T^{3/2}\rho^{-1/2}$.
\par
3. The acceleration in a collapsing system increases as the collapse proceeds
 (after detachment from the Hubble flow). If $\a0$ varies 
at all, it is expected to decrease with cosmic time. So, the effect of
 MOND is expected to decrease with time in a collapsing system.
 (The system would behave as if the fraction of 
fictitious dark matter it harbors decreases with time.)
\par
In default of a theory one can still
attempt to obtain approximate MOND cosmologies--in order to get a hint
of what is expected--by supplementing
nonrelativistic MOND with 
extra assumptions.
 For instance, one might assume that $\a0$ does not vary with
 cosmic time, identifying it with a veritable cosmological
 constant\cite{com}. This is done in ref. \cite{sancosm} where some further
tentative assumptions are made. In such a case one is bound to ask
why it is that this constant $\a0$ is today of the same order as the
variable $c\h0$. The same question arises in connection with the 
emerging value of the cosmological constant $\L\sim \h0^2$.
In MOND, at any rate, this could find an antropic explanation whereby
structure formation (hence star formation and the eventual development
of mankind)
 is facilitated when the acceleration within the horizon 
($\sim c\h0$)--decreasing as it does with cosmic time--becomes
 similar to the crucial dynamical constant $\a0$\cite{com}\cite{sancosm}.

\section{A possible origin of MOND\label{origin}}
Why should then a cosmological acceleration parameter enter local dynamics
in galaxy systems? I have discussed this question in
refs.\cite{ann}\cite{vacuum} and give here a brief account.
I shall concentrate on mondified inertia,
which seems to me more promising at this juncture.
\par  
The thread I would like to follow is that inertia might result from the
 interaction of matter with the vacuum. Also, cosmology affects the
 vacuum and is affected by it (e.g. through a contribution to a cosmological
 constant).
So, either cosmology affects inertia through the intermediary vacuum, 
 or, cosmology and inertia are both affected by the vacuum dynamics, which
then enters cosmology, say, as a cosmological constant, $\L$, and MOND through 
$\a0\approx c\L^{1/2}$.
\par
Inertia is what makes kinematics into dynamics, associating with motion the
attributes of 
energy and momentum that can be changed only by applying forces, as described
by the appropriate equation of motion. Just how much energy and momentum
is associated with so much motion is dictated by the kinetic action of the 
relevant degrees of freedom. To obtain inertia as a derived effect
 is to derive the kinetic actions (in our case from some vacuum effect).
From this action the energy-momentum tensor is derived; thus,
 in relativity, this
action also encapsules the contribution of the
 particular degree of freedom to the
 sources of gravity. Attempts to derive inertia--in
 the spirit of Mach's principle--have concentrated mainly on inertia of
 bodies--see e.g. ref. \cite{bar}. But, of course,
 all dynamical degrees of freedom, whether we describe
them as bodies (particles) or fields, carry inertia.
\par
Supposedly one starts from only interactions between the different degrees
of freedom and get inertia in the form of effective kinetic actions. 
 We know that interactions can, indeed, induce and modify inertial actions.
For example, the effective mass of ``free'' electrons and holes in a 
semiconductor can be greatly changed from its vacuum value; 
 mass renormalization 
in field theory	is, of course, a vacuum effect; and the Higgs mechanism induces
an effective mass term from the interaction with the putative Higgs field.
It is also known that the interaction of the electromagnetic field with
charged vacuum fields begets a free effective action for the 
electromagnetic field--the Heisenberg-Euler effective action
(see e.g. \cite{zel} and \cite{izu} p. 195).
What role, if any, these mechanisms play in MOND is not clear.
However, since they are known to affect inertia, they must be reckoned with in
any complete analysis. 
\par
The scheme I have in mind is inspired by 
  Sakharov's proposal\cite{sakh} to derive the ``free'' (Einstein-Hilbert)
 action of gravity from effects of 
the vacuum: Curvature of space-time modifies the dynamical behavior of
 vacuum fields, hence producing an
associated energy or action for the metric field. To lowest order (in the 
Planck length over the curvature radius) this gives the desired expression
 $\int g^{1/2}R$. Sakharov's 
arguments make use of the fact that the vacuum fields have inertia (since
they are assumed to carry the usual energy-momentum). So, derived inertia 
comes prior to induced gravity a-la Sakharov. Mechanisms proposed in the literature
to produce inertia from vacuum effects (as in refs.  \cite{jr} \cite{rh})
 also presuppose inertia of the vacuum fields, and can thus not serve
 as primary mechanisms for inertia.
\par
For the vacuum to be an agent for inertia
 it is necessary, in the first place, that
a non-inertial observer be able to perceive enough
 details of its motion in the vacuum. The Lorentz invariance
 built into our theories leads to a vacuum that is,
perforce, Lorentz invariant, so uniform motion cannot be detected through it.
It is well known, however, that non-inertial motion raises from the vacuum
a specter that can be sensed by the observer in different ways\cite{unruh}.
 This phenomenon
has so far been studied for only a limited class of simple motions.
 For example, for an observer on a collinear trajectory of
 constant-acceleration, $a$, (hyperbolic motion) this avatar of the vacuum
is the Unruh radiation: a thermal bath the observer finds itself immersed in,
of temperature $T=a/2\pi$ 
($\hbar=1$, $c=1$, $k=1$)\cite{unruh}\cite{birdav}.
 Circular, highly relativistic motions have been discussed, e.g. in refs.
 \cite{ger}-\cite{leinaas} where it is found that a single parameter, 
 $a=\c^2 v^2/r\approx \c^2/r$, still determines the spectrum of
 the incarnation of the vacuum  ($\c$ is the Lorentz factor); this 
 is quasi-thermal with effective temperature $T=\eta a/2\pi$,
  where $\eta$ is of order unity and depends somewhat on the frequency.
 For general motions, hardly anything is known about the radiation. It is
clear that the effect must be a
nonlocal functional of the whole trajectory, because the relevant
 wavelengths and
frequencies of the radiation may be of the order of scale lengths and
 frequencies, respectively, that
characterize the motion.  
(For stationary motions, such as the two described above,
 all points are equivalent, so the Unruh-like
radiation appears to depend only on ``local'' properties. However, the
non-local information on the stationarity of the
 trajectory enters strongly.)	
\par
While the Unruh-like radiation may well serve as a
 marker for non-inertial motions
it is still difficult to implicate it directly in the
 generation of inertia:
1. It is not clear that it carries all the information on the motion 
 needed to produce inertia. For example, even for hyperbolic
 motion, can the 
direction of its acceleration be told by the accelerated
 observer (it should be
 remembered that the radiation is characterized by more
 than just its spectrum.
For example, a finite size observer can compare the radiation
 in its different
 parts.)
2. If inertia is local--as it is to a very good approximation
 in the non-MOND
regime--it has to adjust instantaneously to the state of
motion. The latter may change however on time scales that are
 short compared
 with the typical period of the Unruh-like radiation. In the
 MOND regime there 
is no experimental indication that inertia is local; on the
 contrary, as mentioned
before, theoretical arguments point to nonlocal MOND inertia.
\par
How does MOND fit into this, and, in particular, how can the
 connection with
cosmology be made? 
When the acceleration of a constant-$a$ observer becomes smaller
 than $\a0$,
the typical frequency of its Unruh radiation becomes smaller than
 the expansion
rate of the Universe, the Unruh wavelength becomes larger than
the Hubble distance, etc. \cite{com}\cite{ann}. We expect then
 some break in
 the response
of the vacuum when we cross the $\a0$ barrier. What is the Unruh radiation
seen by a non-inertial observer in a nontrivial universe?
 We know that even inertial observers
in a nontrivial universe find themselves immersed in radiation arising
from the distortion of the vacuum.
 The simplest and best-studied case is that of a de Sitter
 universe in which all
inertial observers see a thermal spectrum with a temperature 
$\tl=(\L/3)^{1/3}/2\pi$ \cite{gibhawk}, where $\L$ is the cosmological
 constant characterizing the de Sitter cosmology. 
It was shown in refs. \cite{npt}\cite{des}
 that an observer on an hyperbolic trajectory, in a de Sitter
 universe, also sees thermal radiation, but with a temperature
\beq T(a)={1\over 2\pi}(a^2+\L/3)^{1/2}.  \label{pultada} \eeq
\par
If inertia is what drives a non-inertial body back to (some nearby)
inertial state, striving to annul the vacuum radiation--here,
 for hyperbolic motion, to drive $T$ back to $\tl$--then
 $T-\tl$ is a relevant quantity. (With cosmology fixed, the best that
 inertia can do is drive $T$ to $\tl$; in the cosmological context it
also strives to drive $\tl$ to zero.)
We can write
\beq  2\pi(T-\tl)\equiv2\pi\Delta T= a\hat\mu(a/\hat\a0), \label{musata}
 \eeq
 with
\beq \hat\m(x)=[1+(2x)^{-2}]^{1/2}-(2x)^{-1},  \label{mura} \eeq
and $\hat\a0=2(\L/3)^{1/2}$.
The quantity $\Delta T$ behaves in just the manner required from MOND 
inertia\cite{mond} [$\hat\mu(x\ll 1)\approx x$,
 $\hat\mu(x\gg 1)\approx 1-(2x)^{-1}$] with $\a0=\hat\a0$
 naturally identified
 with a cosmological acceleration parameter.
(This need not be the effective form of $\mu$ for trajectories
 other than hyperbolic; in mondified inertia there is no $\mu$ in the
 theory itself, and a different form of $\mu$ may apply,
 for instance, to circular orbits\cite{ann}\cite{vacuum}.)
While this observation
 is interesting and suggestive, I cannot tell whether it is germane
to MOND, because it is not backed by a concrete mechanism for inertia, and
because I cannot generalize the observation to more general motions. 
\par
In de Sitter space-time the expansion rate, the space-time
curvature, and the cosmological constant are one and the same.
These parameters differ from each other in a general Friedmanian
universe, and so the above lesson
learnt for the de Sitter case does not tell us which of the cosmological
acceleration parameters is to be identified with $\a0$ in the
 real universe. 
\par
Recall that, in MOND, inertia vanishes in the
 limit $\a0\rar \infty$. In the above
 picture this qualitative tenet of MOND is effected because the limit
 corresponds to $\L\rar\infty$, or 
$\h0\rar \infty$, etc.; so, the Gibbons-Hawking-like radiation
due to cosmology swamps the
thermal effects due to non-inertial motion: the 
difference between inertial
and non-inertial observers is effaced in this limit.

\section*{Acknowledgments}
I thank my friends Jacob Bekenstein and Bob Sanders for many helpful
comments and discussions from which I have benefited over the years.


\vspace{-14pt}
\begin{thebibliography}{99}
\bibitem{mond} Milgrom M 1983  Astrophys. J. {\bf  270} 365
\bibitem{bm} Bekenstein J D  and Milgrom M 1984 Astrophys. J. {\bf 286} 7
\bibitem{ann} Milgrom M 1994  Ann. Phys. {\bf 229} 384
\bibitem{sol} Milgrom M 1986  Astrophys. J. {\bf  302} 617
\bibitem{galaxies} Milgrom M 1983  Astrophys. J. {\bf  270} 371
\bibitem{gerhard} Gerhard O E 1994  {\it Proc. of the ESO/OHP Workshop,  dwarf galaxies}
Eds. Meylan G and Prugniel P (Garching: ESO) p 309 
\bibitem{dwarfs} Milgrom M 1995  Astrophys. J. {\bf 455}  439
\bibitem{mcde} McGaugh S S  and de Blok  W J G  1998 Astrophys. J. {\bf 499} 66
\bibitem{mateo} Mateo M Olszewski E W Vogt S S and Keane M J 1998,  preprint astro-ph/9807296
\bibitem{bbs} Begeman K G Broeils A H and Sanders R H 1991 Mon. Not. R. Astron. Soc. {\bf 249} 523
\bibitem{sanrot} Sanders R H 1996 Astrophys. J., {\bf 473} 117
\bibitem{sanver} Sanders R H and Verheijen  M A W 1998 Astrophys. J. {\bf 503} 97
\bibitem{demc} de Blok  W J G  and McGaugh S S  1998 Astrophys. J.  in press astro-ph/9805120
\bibitem{groups} Milgrom M 1998 Astrophys. J. Lett. {\bf 496}  L89
\bibitem{thewhite} The  L S  and White S D M 1988 Astron. J. {\bf 95} 1642
\bibitem{gerbal} Gerbal D  Durret F   Lachi\'ez-Ray M and Lima-Neto  G 1992 Astron. and  Aastrophys.  {\bf 262} 395
\bibitem{sanclusta} Sanders R H 1994 astron. astrophys. Lett. {\bf 284} L31
\bibitem{sanclustb} Sanders R H 1998 astro-ph/9807023
\bibitem{filaments} Milgrom M 1997  Astrophys. J.  {\bf  478}  7
\bibitem{rome} Milgrom M 1996 {\it Proc. Workshop on The Dark Side of the
 Universe } Eds. Bernabei R Incichitti A (Singapore World scientific) p 2 
\bibitem{mittaz} Mittaz J P D Lieu R Lockman F J 1998 Astrophys. J. Lett. {\bf 498} L17
\bibitem{systems} Milgrom M 1983  Astrophys. J. {\bf  270} 384
\bibitem{kent} Kent S M 1987 astron. J. {\bf 93} 816
\bibitem{morishima} Morishima T and Saio H 1995 astrophys. J. {\bf 450} 70
\bibitem{vacuum} Milgrom M 1998 preprint astro-ph/9805346
\bibitem{stab} Milgrom M 1989  Astrophys. J. {\bf  338} 121
\bibitem{bekrel} Bekenstein J D 1992 in Proc. 6th Marcel Grossman Meeting
 on GR, eds. Sato H and Nakamura T (Singapore: World scientific) p 905
\bibitem{bsbending} Bekenstein J D  and Sanders R H  1994 Astrophys. J. {\bf  429} 480
\bibitem{stratified}Sanders R H  1997 Astrophys. J. {\bf 480} 492
\bibitem{com} Milgrom M 1989 Comments  Astrophys. {\bf 13(4)} 215
\bibitem{sivaram}Sivaram C 1994 Astrophys. Space Sci {\bf 219} 135
\bibitem{sancosm} Sanders  R H  1998 Mon. Not. R. Astron. Soc. {\bf 296} 1009
\bibitem{bar}Barbour J B and Pfister H (eds.) 1995 {\it Mach's Principle}
   (Boston Birkhauser)
\bibitem{zel} Zeldovich Y B 1967 Pis'ma Zh. Eksp. Teoret. Fiz. {\bf 6} 922
(English translation in JETP Lett. {\bf 6} 345 1967) 
\bibitem{izu}  Itzykson C and Zuber J B 1980 {\it Quantum Field Theory} McGraw-Hill
\bibitem{sakh}  Sakharov A D 1968 Sov. Phys. Doklady {\bf 12} 1040 
\bibitem{jr} Jaekel M T and Reynaud S 1993 J. de Physique {\bf 3} 1093
\bibitem{rh}  Rueda A and  Haisch B 1998 Phys. Lett. A {\bf 240} 115
\bibitem{unruh} Unruh W G 1975 Phys. Rev. D {\bf 14} 870
\bibitem{birdav} Birell N D and Davies P C W 1982, {\it Quantum Fields in
Curved Space} Cambridge University Press (Cambridge)
\bibitem{ger} Gerlach U H 1983 Phys. Rev. D {\bf 27} 2310
\bibitem{bell} Bell J S and Leinaas J M 1987 Nuc. Phys. {\bf B284} 488
\bibitem{lpp} Levin O Peleg Y and Peres A 1993 J. Phys. A: Math Gen. {\bf 26} 3001
\bibitem{audretsch} Audretsch J M\"uller R and Holzmann M 1995 Class.
 Quantum Grav. 12 2927
\bibitem{untwo} Unruh W G 1998 preprint hep-th/9804158
\bibitem{leinaas}  Leinaas J M 1998 preprint hep-th/9804179
\bibitem{gibhawk}  Gibbons G W and Hawking S W 1977 Phys. Rev. D {\bf 15} 2738
\bibitem{npt} Narnhofer H Peter I and Thirring W 1996, Int. J. Mod.
 Phys. B {\bf 10} 1507
\bibitem{des}  Deser S and Levin O 1997 Class. Quant. Grav. {\bf 14} L163

\end{thebibliography}
\end{document}